\documentclass[%
 aip,
 amsmath,amssymb,
preprint,%
]{revtex4-1}

\usepackage{graphicx}
\usepackage{dcolumn}
\usepackage{bm}

\usepackage[utf8]{inputenc}
\usepackage[T1]{fontenc}
\usepackage{mathptmx}
\usepackage{float}

\begin{document}

\preprint{}

\title{Elastocaloric effect in amorphous polymer networks undergoing mechanotropic phase transitions}
\author{J.\,A.~Koch}
\author{J.\,A.~Herman}
\affiliation{University of Colorado Boulder, Department of Chemical and Biological Engineering}

\author{T.\,J.~White}
\email{timothy.j.white@colorado.edu}
\affiliation{University of Colorado Boulder, Department of Chemical and Biological Engineering}
\affiliation{University of Colorado Boulder, Materials Science and Engineering Program}

\date{1 March 2021}

\begin{abstract}
Deformations of amorphous polymer networks prepared with significant concentrations of liquid crystalline mesogens have been recently reported to undergo mechanotropic phase transitions. Here, we report that these mechanotropic phase transitions are accompanied by an elastocaloric response ($\Delta T = 2.9$~K). Applied uniaxial strain to the elastomeric polymer network transitions the organization of the material from a disordered, amorphous state (order parameter $Q=0$) to the nematic phase ($Q=0.47$). Both the magnitude of the elastocaloric temperature change and mechanically induced order parameter are dependent on the concentration of liquid crystal mesogens in the material.  While the observed temperature changes in these materials are smaller than those observed in shape memory alloys, the responsivity, defined as the temperature change divided by the input stress, is larger by an order of magnitude. 
\end{abstract}

\maketitle

\section{Introduction}

Liquid crystal mesogens populating an amorphous polymeric network may develop nematic order upon applied deformation. This ``mechanotropic'' phase change was the subject of a recent work,\cite{donovan2019mechanotropic} which characterized the order parameter ($Q$) with experimental measurements demonstrating an increase from $Q=0$ at uniaxial strain $\epsilon = 0\%$ to $Q>0.4$ at $\epsilon=225\%$. We report here that this phase change is accompanied by an elastocaloric temperature change.

Elastocaloric effects in materials are associated with a coupling of macroscopic deformation and microstructural rearrangement. An applied deformation may increase or decrease the configurational entropy of a material by, for example, inducing a crystalline/crystalline transition\cite{bonnot2008elastocaloric,tuvsek2015elastocaloric} or an disordered-to-ordered phase transformation\cite{guyomar2013elastocaloric}. If the deformation is an adiabatic process, this change in configurational entropy will be compensated by a change in ``thermal'' entropy -- i.e., a change in temperature\cite{xie2016comparison,ikeda2019ac} (see Supplementary Material for a more detailed derivation). In other words: a deformation that increases microstructural order, thereby decreasing configurational entropy, will result in a temperature increase ($\Delta T^+$).  Reversal of this process (e.g., by releasing the stress applied to the material) enables the microstructure to return to a less ordered (disordered) state and results in a temperature decrease ($\Delta T^-$). Elastocaloric phenomenon are but one of several solid-state caloric effects, notably including electro, magneto, and barocaloric phenomena, that are currently being explored for functional implementation in next-generation refrigeration systems.\cite{moya2014caloric,moya2020caloric}

Elastocaloric effects have been described since at least 1802, when the philosopher John Gough used his lips to sense the heating of natural rubber as it was rapidly stretched.\cite{gough1806} In part motivated by a 2014 U.S. Department of Energy report,\cite{goetzler2014energy} elastocaloric heat engines are being considered as candidates to replace vapor-compression refrigeration systems. This report highlighted shape memory alloys (SMAs) as the pacesetter for this technology. Most notable within this class of materials is Nitinol, filaments of which have demonstrated an adiabatic temperature increase of $\Delta T^+=25.5$~K upon application of a strain of $\epsilon=8.5\%$.\cite{cui2012demonstration} A recent report documents an elastocaloric response of 31.5~K.\cite{cong2019colossal} While these mechanically induced temperature changes are impressive, the stress required to cause the associated phase transition are on the order of $\sigma=500-700$~MPa, making the elastocaloric ``responsivity'' of SMAs (the temperature increase normalized by the applied stress, $\Delta T/\sigma$) approximately 0.045~K/MPa.

Conversely, while the relative magnitude of elastocaloric temperature changes observed in soft materials is smaller ($\Delta T=$~0.4~to 10~K), these materials can exhibit responsivity values of up to 10~K/MPa.\cite{xie2017temperature,trvcek2016electrocaloric}  Further, soft materials may have functional benefits in small-scale implementations due to the order-of-magnitude reduction in required stress, which could reduce the corresponding power input, cost, and weight of the auxiliary equipment. The comparative increase in the rate of elastic recovery of soft materials could enable higher frequency load/unload cycles in end-use implementations.

Natural rubber continues to be a subject of elastocaloric research. Recent work has characterized a mechanically induced temperature increase of $\Delta T=2$~K in rubber, at strains in excess of $\epsilon>500\%$.\cite{guyomar2013elastocaloric} Since a majority of this deformation does not contribute to substantial microstructural rearrangement, prestrain of the material can be utilized to limit the amount of elongation needed to develop strain-induced crystallization, thereby achieving greater temperature change, $\Delta T^+=4.3$~K, for applied $\epsilon<100\%$.\cite{xie2015elastocaloric} 

Elastocaloric effects in polymeric materials have been examined in other geometries and materials systems. For example, a recent study details sizable elastocaloric output in coiled elastomeric fibers (a ``twistocaloric'' effect). The outer surface of the coiled natural rubber achieves $\Delta T^+>10$~K at material strains of $\epsilon=600\%$ with no axial elongation.\cite{wang2019torsional} Another recent report details the construction of an inflatable natural rubber membrane that combines strain-induced crystallization and a snap-through instability to generate an elastocaloric $\Delta T^+=$7.9~K.\cite{greibich2021elastocaloric} The nascent processability of polymers evident in these works are an emerging approach to further enhance elastocaloric performance ($\Delta T$, responsivity) in these materials. Other polymeric systems, such as poly(cyclooctene), have been examined with an elastocaloric response of $\Delta T^+=2.8$~K at 300\% strain.\cite{hong2019characterization} Polyvinylidene difluoride (PVDF) and related co- and terpolymers are notable multi-caloric materials. A 1.8~K elastocaloric temperature change was observed in PVDF across a broad range of temperatures (300-350~K) when subject to an applied stress of 15~MPa.\cite{patel2016elastocaloric} In poly(vinylidene fluoride-trifluoroethylene-chlorofluoroethylene), a slighly larger temperature change (2.2~K) was observed when this material was subject to an applied stress of 150~MPa.\cite{yoshida2016elastocaloric} Comparatively, PVDF exhibits a barocaloric temperature change of up to 19~K at an elevated ambient temperature\cite{patel2016elastocaloric} and an electrocaloric temperature change of up to 11~K.\cite{lu2018enhancing}

Here, we are concerned with the elastocaloric response of liquid crystal-containing elastomers. Notably, these materials are relatives, but also distinguished from, so-called liquid crystalline elastomers (LCEs) in that they originate from an amorphous state (versus a polydomain or aligned state). Liquid crystal (LC) mesogens, typically made up of multiple aromatic rings that introduce strong intramolecular $\pi-\pi$ coupling, organize most commonly into nematic (orientational order) or smectic (orientational, positional order) phases. When incorporated into an elastomeric polymeric network, the mesogenic segments participate in reversible elastic processes. 

\begin{figure}
\centering
\includegraphics[width=.58\textwidth]{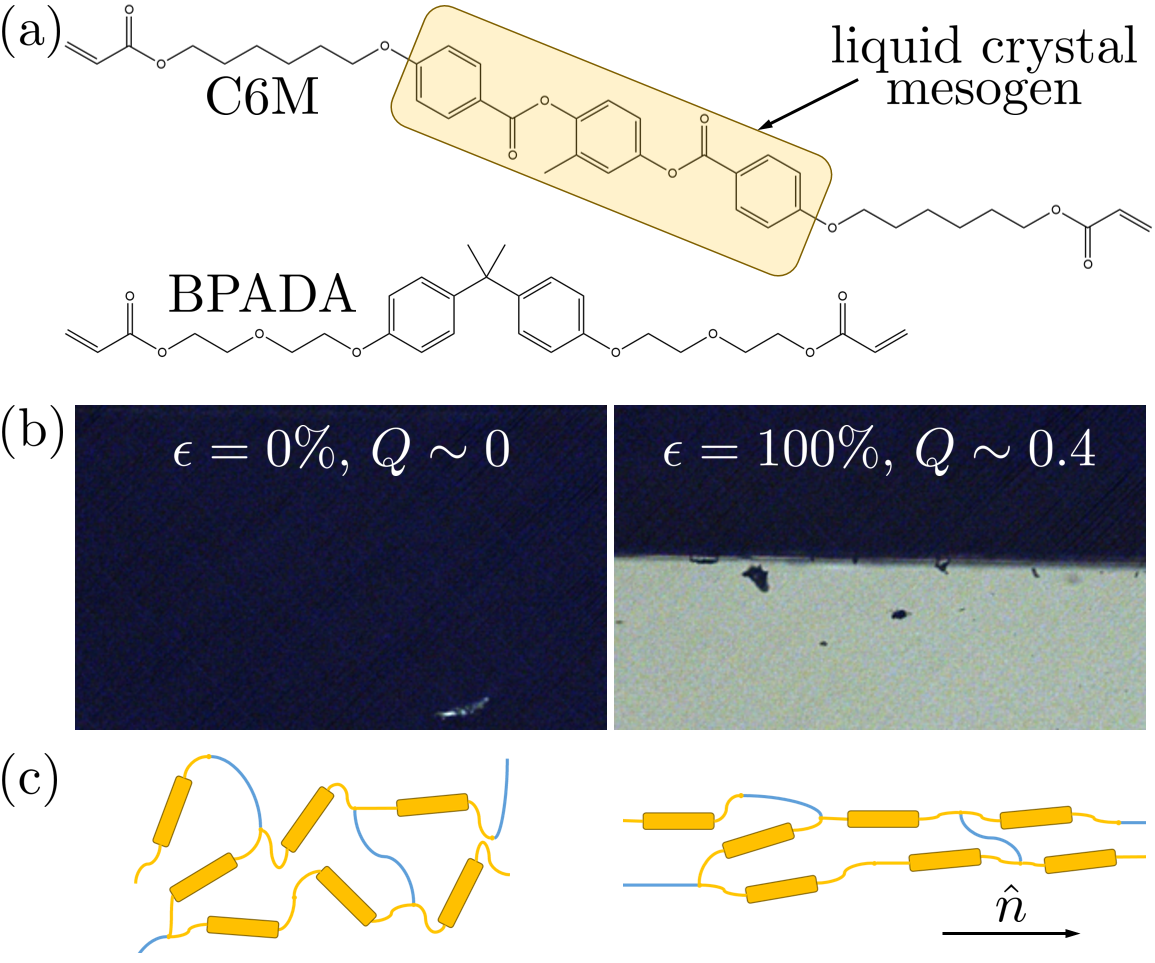}
\caption{\label{fig:network} (a) The polymeric network is made up of a diacrylate main-chain liquid crystal monomer, C6M (1,4-bis-[4-(6-acryloyloxyhexyloxy)benzoyloxy]-2-methylbenzene), and a non-liquid crystalline diacrylate, BPADA (bisphenol A ethoxylate diacrylate) prepared via thiol-acrylate photopolymerization. (b) The elastomer becomes birefringent as strain is applied due to the development of nematic alignment upon elongation. (c) The network architecture, made up of liquid crystalline mesogens (yellow cylinders) and non-liquid crystalline monomers (blue), is amorphous at room temperature. Deformation to load aligns the network, a ``mechanotropic'' phase change, causing the development of nematic order (defined by the director $\hat{n}$) among the mesogens along the loading direction. }
\end{figure}

Recent literature reports detail the elastocaloric response of LCEs. Experimental efforts have so far demonstrated elastocaloric temperature changes of only $\Delta T\sim 1$~K. \cite{trvcek2016electrocaloric,lavrivc2020tunability} Lavrič et~al.~observed $\Delta T = 1$~K at a strain of $\epsilon=90\%$ and detailed that this $\Delta T$ was observed across a range of crosslink density.\cite{lavrivc2020tunability}
Skačej has reported molecular simulations that suggest the elastocaloric effect in main-chain LCEs could yield a $\Delta T>10$~K.\cite{skavcej2018elastocaloric} However, the simulations underestimate the specific heat of liquid crystal-containing systems, which likely leads to an overestimate of the elastocaloric temperature change. LC systems also have been explored as electrocaloric materials, notably harnessing a smectic-to-isotropic transition to achieve a $\Delta T^+$ of up to 5.2 to 6.5~K.\cite{trvcek2016electrocaloric,klemenvcivc2019giant}

Our group has recently quantified the transition from disorder (order parameter $Q\approx 0$) to appreciable order ($Q\approx 0.4$, i.e., the nematic phase) in a liquid crystal-containing elastomeric network upon application of uniaxial strain.\cite{donovan2019mechanotropic} This ``mechanotropic'' phase change, from amorphous to liquid crystalline (see footnote\footnote{A distinction is made between the ``amorphous'' liquid crystal-containing elastomers in the present work and ``isotropic'' liquid crystal elastomers. The latter implies the existence of a thermally induced phase change to a liquid crystal nematic upon cooling. Such a phase change is not generally seen in the materials described here due to a comparatively lower concentration of liquid crystal mesogens in the network architecture.} on ``amorphous'' vs.~``isotropic''), has unexplored potential relevance as an elastocaloric material system in that applied mechanical force can induce microstructural order that, if adiabatic, would be accompanied by a change in temperature.

We report here a stretch-induced temperature increase of up to $\Delta T=3$~K in these liquid crystal-containing elastomers, observed at a strain of $\epsilon=100\%$ and a stress of $\sigma\approx 2$~MPa (a responsivity of 1.5~K/MPa). The nematic order parameter, defined as
\begin{equation}
    Q=\langle \tfrac{3}{2}\cos^2\theta -\tfrac{1}{2}\rangle
\end{equation}
(an averaging of the orientation $\theta$ of the LC director, $\hat{n}$)
is calculated directly by measuring the polarized absorbance of the polymer network doped with a dichroic molecule. Order increases with the magnitude of strain and liquid crystal concentration.  These trends are mirrored in the associated increase in elastocaloric temperature change.  Accordingly, we conclude the observed temperature increase is associated with mechanically induced order. That is, the mechanotropic phase transition is elastocaloric in nature.

\section{Materials and methods}

The polymeric networks studied in this work were prepared in a one-pot synthesis primarily based on the copolymerization of two diacrylate monomers: the liquid crystal monomer 1,4-bis-[4-(6-acryloyloxyhexyloxy)benzoyloxy]-2-methylbenzene (Wilshire Technologies, C6M) and the non-liquid-crystalline monomer bisphenol A ethoxylate diacrylate (Aldrich, BPADA). Chemical structures are shown in Fig.~\ref{fig:network}(a). Elastomeric polymer networks were prepared from mixtures of C6M and BPADA monomers that were first oligomerized by a thiol-Michael addition reaction.  Subsequently, the acrylate end-capped oligomers were photopolymerized to produce a fully-cured, crosslinked network. After harvesting, the materials were cut into strips with dimension 5-by-1.5-by-0.1 mm$^3$. Additional details of the synthesis are included in Supplementary Materials.

Similar to the disruption of liquid crystalline phases by the inclusion of non-mesogenic organic solvents, basing the formulations on appreciable concentration of BPADA disrupts the nascent liquid crystallinity of C6M so that the polymerization of these materials originates from an isotropic state.  Upon polymerization from this state, these materials form amorphous polymer networks that are populated with randomly oriented liquid crystalline segments (yellow cylinders, Fig.~\ref{fig:network}(c)). The liquid crystal concentrations examined here vary between 0 and 60~wt\% C6M. The application of uniaxial strain aligns the chain orientation in the network, allowing the randomly aligned liquid crystal mesogens to organize into nematic phase that is evident in a large birefringence (Fig.~\ref{fig:network}(b)) and order parameter $Q$ exceeding 0.4.\cite{donovan2019mechanotropic}

Strain was imposed and stress measured using a TA~Instruments dynamic mechanical analyzer (Discovery DMA~850). ``Strain'' throughout refers to engineering strain, $\epsilon\equiv (L-L_0)/L_0$, for stretched length $L$ and initial length $L_0$. Temperature changes during deformation and recovery were collected from infrared video (FLIR~T630sc recording at 30~fps). Temperatures were averaged over a rectangular region encompassing the middle third of the sample length and width.  The sampling frame was adjusted as the elastomer was strained.

Order measurements were undertaken by doping the material with 3~wt\% 4,4'-bis(6-acryloyloxy hexyloxy)azobenzene (Synthon, referred to hereafter as azobenzene), a dichroic molecule which incorporates into the polymeric network.  Azobenzene has a strong absorbance peak at $\lambda=365$~nm. As the elastomer is stretched, the azobenzene is reoriented concurrent to the the polymer strands in the network.  Accordingly, the dichroic absorption of azobenzene is introduced and can be used as a means of quantifying orientation. Time-dependent transmission measurements were made using an STS-VIS spectrometer (Ocean Insight). Steady-state measurements of the materials at designated deformations were carried out on a Cary 7000 UV-VIS-NIR with a linearly polarized sample beam. The order parameter is calculated by comparing the absorbance of light polarized parallel ($A_\parallel$) and perpendicular ($A_\perp$) to the direction of elongation according to\cite{rumi2016quantification}
\begin{equation}\label{GHeqn}
    Q=\frac{A_\parallel-A_\perp}{A_\parallel+2A_\perp}.
\end{equation}
An example of this absorbance data is available in the Supplementary Material, Fig.~S1.

\section{Results}

\subsection{Temperature change measurements}
\begin{figure*}
\centering
\includegraphics[width=0.98\textwidth]{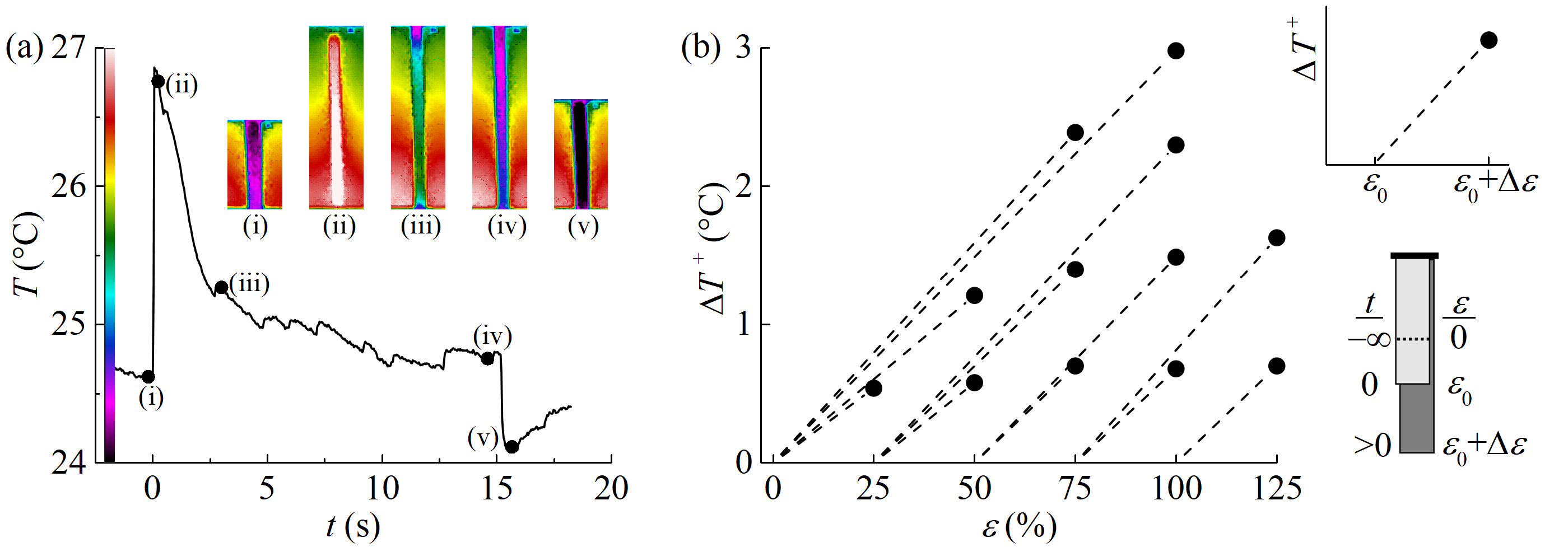}
\caption{\label{fig:T_profile} (a) A typical experiment to measure the elastocaloric effect. The elastomer (here, prepared with 60~wt\%~C6M) begins at room temperature with no external strain. At $t=0$, a step strain to $\epsilon=\epsilon_0$ (here, $100\%$) is applied,
causing an increase in the material temperature of $\Delta T^+=2.3$~K. The sample remains stretched but loses heat to the environment, returning to room temperature. At $t=15$~s the strain is released, and the sample cools by $\Delta T^- =-0.7$~K. Inset images (i)-(v) are infrared stills, taken at times denoted on the temperature-time curve. In (b) the roles of prestrain $\epsilon_0$ and step-strain magnitude $\Delta\epsilon$ are examined. The material, beginning with no strain (at $t=-\infty$), is slowly stretched to a designated strain value $\epsilon_0$ before an step-strain of $\Delta\epsilon$ is applied at $t=0$, for a total material strain of $\epsilon=\epsilon_0+\Delta\epsilon$. A similar elastocaloric temperature change occurs regardless of prestrain value $\epsilon_0$ ($x$-intercept of the dotted lines), with the temperature increase being proportional to the strain step size $\Delta \epsilon$ (change in $x$-coordinate of the dotted line). The slopes of the dotted lines, $d(\Delta T)/d\epsilon$, are approximately the same regardless of the prestrain value $\epsilon_0$, indicating the eC temperature change is proportional to step-strain size: $\Delta T^+/\Delta \epsilon = 2.6\pm0.3$~K/(-).}
\end{figure*}

Amorphous polymer networks prepared with appreciable liquid crystalline content undergo mechanotropic phase transitions to load from the disordered to nematic phase. We hypothesize that the associated decrease in configurational entropy will be compensated by an increase in temperature (assuming the process is adiabatic). In Fig.~\ref{fig:T_profile} a step strain of $\Delta\epsilon=100\%$ is applied at $t=0$ to a material containing 60~wt\%~C6M. The elastomer responds with an instantaneous temperature increase, from $T=24.6$ to 26.9$^{\circ}$C ($\Delta T^+=2.3$~K). As the strain is held, the material loses heat to the environment and cools back to room temperature. At $t=15$~s, the strain is released, and the material drops in temperature from $T=24.8$ to 24.1$^{\circ}$C, below room temperature ($\Delta T^-=-0.7$~K). Subsequently, the material warms back to room temperature.

The step-strain deformation in Fig.~\ref{fig:T_profile} occurs over 0.048~s while the temperature change in the material is observed over 0.067~s (two video frames), indicating the temperature change is coupled to the elongation. The ensuing temperature decay has a thermal time constant of approximately 1.9~s -- much slower than the imposed strain input and observed temperature jump.  Accordingly, we conclude the elongation process is adiabatic. The presence of a jump decrease in temperature $\Delta T^-$ upon strain release is further evidence the mechanically induced temperature change is not directly related to mechanical work. We therefore conclude that the observed mechanically induced changes in temperature are attributable to an elastocaloric effect in the material.

Figure~\ref{fig:T_profile}(b) examines the role of prestrain and step-strain magnitude on the elastocaloric temperature increase for the elastomer containing 60~wt\% LC content. The materials were slowly stretched (over 1~minute) to a pre-selected strain value $\epsilon_0$. Subsequently, a step-strain of $\Delta \epsilon$ was applied. Dotted lines are drawn as a guide from the prestrain value ($x$-intercept) to a coordinate pair corresponding to the final material strain, $\epsilon=\epsilon_0+\Delta\epsilon$, and the measured temperature increase $\Delta T^+$. The temperature increase is insensitive to the prestrain value, with $\Delta T^+$ similar regardless of the prestrain amount for a given step-strain magnitude. This is a notable difference from natural rubber, where prestrain places the material near the onset of strain-induced crystallization to achieve greater temperature change for smaller strain inputs.\cite{xie2015elastocaloric} Figure~\ref{fig:T_profile}(b) further demonstrates the role of strain step-size on the elastocaloric effect. In these materials, the observed temperature changes scale proportionately with the size of the input step-strain, such that $\Delta T^+/\Delta \epsilon = 2.6\pm0.3$~K/(-). That is, a larger elastocaloric effect is seen for a larger strain. We note that the polyacrylate networks that are the basis of this examination limit the recovery rate and are prone to failure for strains beyond $\epsilon\approx 125\%$.

\subsection{Order measurements}
\begin{figure*}
    \centering
    \includegraphics[width=0.98\textwidth]{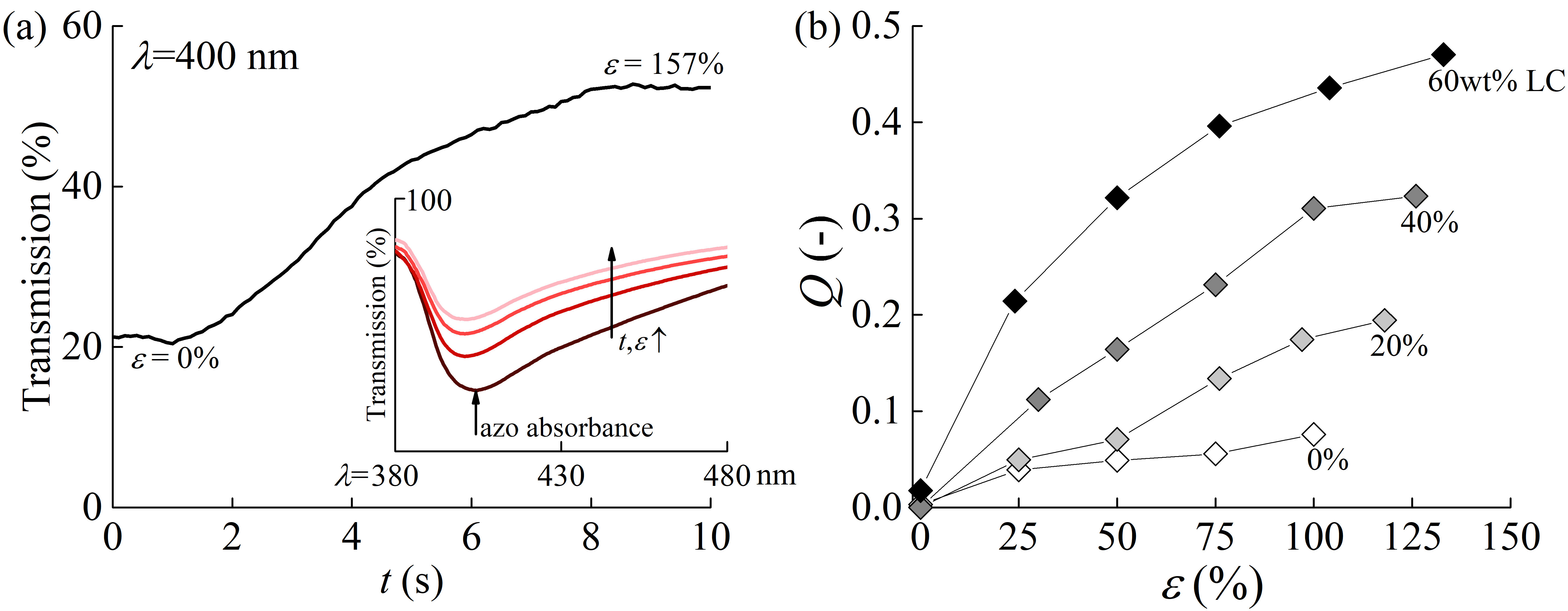}
    \caption{\label{fig:azo_stuff} The liquid crystal-containing elastomers are doped with 3~wt\% azobenzene, a dichroic molecule, and are exposed to polarized light. (a)~For light polarized perpendicular to the direction of elongation, the transmission at $\lambda=$400~nm increases with deformation. This indicates the azobenzene units are aligning to the stretch direction. Inset: the transmission spectrum measurements at $t=2$, 4, 6, and 8~s, demonstrating decreased absorbance as the material becomes stretched. The elastomer stretches from $\epsilon=0$ to 157\% at a rate of approximately $d\epsilon/dt=22\%/\text{s}^{-1}$.
    (b) The order parameter can be calculated for static samples with absorbance measurements perpendicular and parallel to the direction of elongation via Eq.~\eqref{GHeqn}. The order parameter of the materials increases with increasing strain. Further, the concentration of liquid crystalline units in the amorphous elastomers dramatically increases the achievable order parameter by a factor of nearly 6.  This suggests the strain-imposed alignment encourages additional intramolecular interactions among the liquid crystalline segments.}
\end{figure*}

The application of uniaxial strain is hypothesized to cause a nematic alignment of the isotropic distribution of liquid crystal segments in the elastomeric network.  This is evident in the large birefringence evident in Fig.~\ref{fig:network}(b). To further elucidate the dynamics of the mechanotropic phase transitions, we incorporate azobenzene into these compositions and use spectroscopy to measure changes in transmission. In Fig.~\ref{fig:azo_stuff}(a), the transmission at $\lambda=400$~nm is monitored as an elastomer doped with the dichroic absorber azobenzene is uniaxially stretched with strain rate of $d\epsilon/dt= 22$\%/s. The input light source is linearly polarized perpendicular to the direction of stretch. Any preferential alignment of azobenzene to the loading axis will result in an increase in transmitted light (associated with a decrease in absorption) which is evident in the inset to Fig.~\ref{fig:azo_stuff}(a).  

The data in Fig.~\ref{fig:azo_stuff}(a) indicates that there is a direct correlation between strain and orientational order in these materials. The degree to which the liquid crystal content is developing nematic order with strain can be quantified with a series of steady-state absorbance measurements for light polarized parallel and perpendicular to the direction of elongation. In Fig.~\ref{fig:azo_stuff}(b), the order parameter of these materials is estimated by Eq.~\eqref{GHeqn} as a function of applied strain for elastomers containing a range of liquid crystal concentrations. The order parameter increases with increasing strain.  This is a manifestation of a competition between configurational entropy, which has a tendency to drive the polymer network towards an isotropic state, and the applied uniaxial elongation, which aligns the polymer strands (and mesogens within them) to the direction of elongation. This mechanism explains the increase in order upon stretching seen even in the amorphous materials without or with low concentrations of LC mesogens and is corroborated to the elastocaloric temperature change as an adiabatic stretching event.

More significantly, the achievable order increases with increasing LC content, demonstrating that the intramolecular attraction of the mesogenic units in the polymer network are enhancing the system ordering beyond that imposed by the applied strain field. Most notably, the inclusion of 60~wt\% LC monomer enhances the order parameter by nearly a factor of 6 at $\epsilon=100\%$ in comparison to a completely amorphous (0~wt\% C6M) elastomer.

\section{Discussion}
\begin{figure}
\centering
\includegraphics[width=0.58\textwidth]{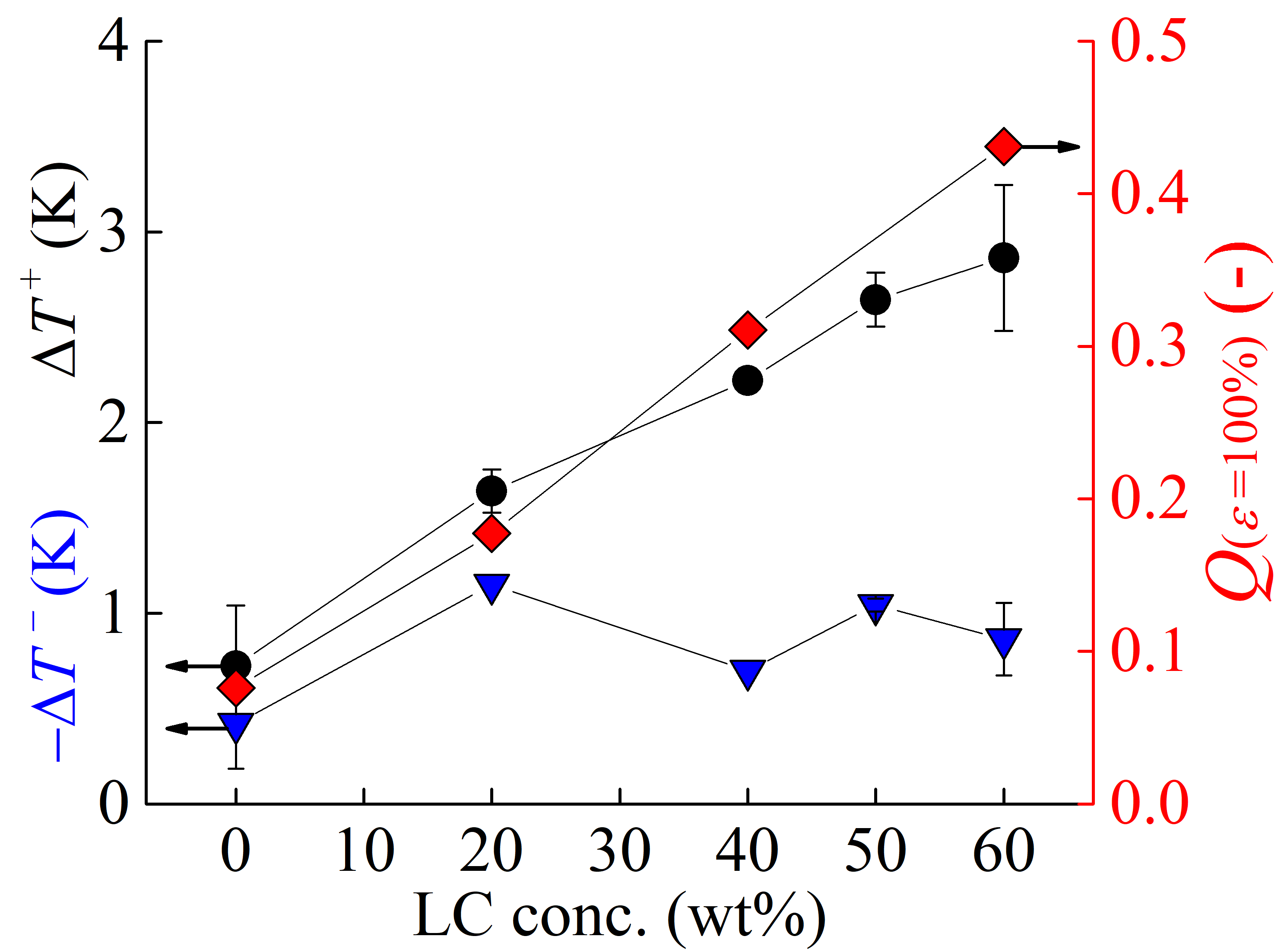}
\caption{The temperature increase upon stretching, $\Delta T^+$, and order parameter, $Q$, increase with liquid crystal content, up to $\Delta T^+=2.9$~K and $Q=0.43$ for materials with 60~wt\% LC content. The temperature decrease upon relaxation, however, is approximately constant as a function of C6M concentration, $\left|\Delta T^-\right|\approx 0.84$~K.}
\label{fig:T_Q_wtp}
\end{figure}
The temperature increase upon strain application, $\Delta T^+$, temperature decrease upon strain release, $-\Delta T^-$, and order parameter, $Q$, are summarized as a function of LC concentration in Fig.~\ref{fig:T_Q_wtp} for an input step-strain of {$\Delta\epsilon=100\%$} (no prestrain). The temperature increase and increase in order parameter associated with deformation increase linearly with increasing concentration of liquid crystalline precursor.  The magnitude of these values increase from $\Delta T^+=0.32$~K and $Q=0.076$ for the purely amorphous elastomer to $\Delta T^+=2.9$~K and $Q=0.43$ for the material prepared with 60~wt\% LC content. Figures~\ref{fig:T_profile}(b) and \ref{fig:azo_stuff}(b) demonstrate that both the elastocaloric temperature change and order parameter increase monotonically with strain.  This indicates the mechanotropic phase transitions examined here are second-order. The parallel influences of liquid crystalline content and mechanical deformation on order and temperature change is convincing evidence that the mechanotropic phase transition in these materials is elastocaloric.

Figure~\ref{fig:T_Q_wtp} also illustrates that the temperature decrease upon strain release does not increase with LC content and remains constant at $|\Delta T^-|=0.84 \pm 0.29$~K. For the amorphous elastomer prepared with 60~wt\% LC content, the temperature increase is larger by a factor $|\Delta T^+/\Delta T^-|=3.3$. Polyacrylate polymer networks notoriously have wide variance in molecular weight between crosslinks (e.g., local crosslink density), potentially leading to long elastic recovery time scales (see Supplementary Material Fig.~S2). Further, the materials chemistry utilized here may also have some degree of branching, which could also hinder the rate of recovery.\cite{jiang2012radical}

We speculate that the forced elongation of the elastomer aligns the LC mesogens rapidly (thus generating a larger $\Delta T^+$) while the relaxation upon strain release occurs over a longer time scale (resulting in a smaller $\Delta T^-$).

The liquid crystal-containing elastomers studied in this work undergo a mechanotropic phase transition, from disordered to nematic, which is asssociated with an elastocaloric temperature change of up to $\Delta T^+=2.9$~K. This represents a large improvement over existing reports on the elastocaloric response in other liquid crystalline polymer networks, which typically demonstrate $\Delta T^+\sim 1$~K,\cite{trvcek2016electrocaloric,lavrivc2020tunability}.  Further, these initial results approach the magnitude of the temperature changes of 5.2 and 6.5~K reported for electrocaloric LC systems.\cite{trvcek2016electrocaloric,klemenvcivc2019giant} While not achieving the elastocaloric temperature change observed in shape-memory alloys such as Nitinol,\cite{cui2012demonstration} the LC-containing elastomers in this work are orders-of-magnitude softer and more responsive ($\Delta T/\sigma$). These soft systems  have nascent benefits in regards to ease of implementation and flexibility in engine design. Ongoing work is exploring other contributing factors to the elastocaloric effect in these materials including polymer network architecture as well as other liquid crystal mesogens.

\section{Conclusions}

The mechanotropic phase change in elastomers with appreciable liquid crystal content is accompanied by elastocaloric output. An elastocaloric temperature jump of $\Delta T^+=2.9$~K was measured for a polymer network populated by 60~wt\% liquid crystal monomer.  This temperature change is associated with an increase in order from $Q\approx 0$ to $Q=0.43$. The magnitude of the elastocaloric temperature change and the order parameter both increase monotonically with the concentration of liquid crystal mesogens in the materials as well as with the magnitude of step-strain. Because the mechanotropic phase transitions in these materials seem to immediately transition into an ordered state, prestrain of these materials does not further enhance the elastocaloric output.  Ongoing research is focused on improving the robustness of the elastomeric materials and exploring the role of mesogen composition on elastocaloric outputs.

\begin{acknowledgments}
We acknowledge financial support from the University of Colorado Boulder. The authors are grateful for an inspired discussion with Zoey S.~Davidson and for experimental assistance from Tayler S.~Hebner and Brian P.~Radka.
\end{acknowledgments}

\nocite{*}
\bibliography{_bib}

\end{document}


\centering
{\Large Supplementary Material}
\vspace{0.5cm}

{\Large Elastocaloric effect in amorphous polymer networks undergoing mechanotropic phase transitions}
\vspace{0.5cm}

J.A.~Koch$^{1}$, J.A.~Herman$^{1}$, T.J.~White$^{1,2}$

$^{1}$University of Colorado Boulder, Department of Chemical and Biological Engineering \\
$^{2}$University of Colorado Boulder, Materials Science and Engineering Program

\raggedright

\appendix
\section{Synthesis details}
The polymeric networks studied in this work were prepared in a one-pot synthesis with a two diacrylate monomers: the liquid crystal 1,4-bis-[4-(6-acryloyloxyhexyloxy)benzoyloxy]-2-methylbenzene (Wilshire Technologies), referred to as C6M, and the non-liquid-crystalline Bisphenol A ethoxylate diacrylate (Aldrich), referred to as BPADA. The C6M and BPADA monomers, along with 2~wt\% photoinitiator (Omnirad~819, IGM~Resins) and 0.5-1~wt\% thermal inhibitor (4-methoxyphenol, Aldrich) were mixed in a vial and melted. Hexane dithiol (Sigma-Aldrich) at a ratio of 0.85~thiol-to-acrylate group was added and the mixture was subject to vortex mixing.  Subsequently, 1~wt\% of base catalyst dipropylamine (Aldrich) was added. The mixture was introduced via pipette to an uncoated glass slide. Addition of a second uncoated glass slide separated by 0.1~mm spacers pressed the drop into a film.  The oligomerization reaction occured on hotplate at $80^{\circ}$C for 3~hours via thiol-addition reaction. The material was then photopolymerized with 405~nm light for 5~minutes per side and allowed to cool before being cut into strips of typical dimension 5-by-1.5-by-0.1 mm$^3$.

\section{Absorbance measurements}
To enable examination of mechanically induced order, the elastomeric networks were doped with 3~wt\% 4,4'-bis(6-acryloyloxyhexyloxy)azobenzene.  The acrylate azobenzene monomer is incorporated into the polymer network via copolymerization.  Azobenzene is dichroic and has a high optical absorption at $\lambda=365$~nm. The experimental data inputs to the calculation of order parameter are presented in Fig.~\ref{fig:abs}. Here, the absorbance is measured with light polarized parallel or perpendicular to the direction of strain, for strains of $\epsilon =$~ 0, 24, 50, and 104\%. For $\epsilon =$0\%, the absorbance curves are similar, though with a slightly higher peak for the parallel data, attributable to a slight strain applied to the material as it was loaded. As the strain increases, the absorbance increases for the light polarized in the direction of the strain, and the absorbance decreases for the light polarized perpendicular to the strain, indicating that the azobenzene units have aligned to the direction of elongation. The absorbance measurements are used to calculate the order parameter ($Q$)~\cite{rumi2016quantification} via
\begin{equation}\label{GHeqn}
    Q=\frac{A_\parallel-A_\perp}{A_\parallel+2A_\perp}.
\end{equation}

\section{Stress measurements}

Figure~\ref{fig:stress} shows an example of the stress response in the liquid crystal-containing materials of this work. In response to a step-strain input (red, presented in detail to demonstrate the time the ``step'' takes to ramp to $\epsilon=100\%$), the stress response (black) typically demonstrates an overshoot followed by a stress decay lasting several seconds. This transient response is likely attributable to the nature of the polyacrylate polymer network and is indicative of long-duration relaxation processes in the network that potentially can hinder the disorder-order and/or order-disorder transition (i.e., as the input strain is applied and/or released). The detailed mechanics of these materials is a subject of future work.

\section{Theoretical rationalization of the elastocaloric effect}
The elastocaloric effect has been rationalized from thermodynamic principles in numerous works, for example in Refs.~\cite{xie2016comparison,ikeda2019ac}. We briefly comment on the derivation here with details relevant to mechanotropic materials.

~\\

The differential free energy $dA$ of the system can be written as
\begin{equation}
    dA = V_0 \sigma d\varepsilon - SdT
\end{equation}
for initial material volume $V_0$, applied stress $\sigma$, differential strain $d\epsilon$, entropy $S$, and temperature differential $dT$.
Additionally, the state can be defined using two variables. Choosing strain $\epsilon$ and temperature $T$, we write
\begin{equation} \label{free_eng}
    dA = \left(\frac{\partial A}{\partial \epsilon}\right)_T d\epsilon +
        \left(\frac{\partial A}{\partial T}\right)_\varepsilon dT.
\end{equation}
From symmetry of second derivatives, we recognize
\begin{equation}
    V_0\left(\frac{\partial \sigma}{\partial T }\right)_\varepsilon=-\left(\frac{\partial S}{\partial \varepsilon }\right)_T
\end{equation}
which can be rearranged and integrated, presuming a straining process that takes the sample from $\epsilon=0$ to $\epsilon=\epsilon_0$, to find
\begin{equation}\label{DeltaS}
    \left(\Delta s\right)_T = -v_0\int_0^{\varepsilon_0} \left(\frac{\partial \sigma}{\partial T }\right)_\varepsilon d\varepsilon
\end{equation}
where the $T$-subscript acknowledges that the derivative of entropy was taken at constant temperature, and we have divided through by the material mass to arrive at intrinsic variables, $\Delta s$ and $v_0$. This equation represents the elastocaloric effect in the limit of constant temperature. In regards to the polymeric materials in this work: To maintain a given strain for an increasing temperature, an increasing stress is in general necessary. The integrand of Eq.~\eqref{DeltaS} is therefore positive, and so $\Delta s<0$. That is, for a constant temperature application of strain $0\rightarrow\epsilon_0$, the system entropy decreases (i.e.,~the microstructure becomes more ordered).

~\\

We can also write the state of entropy in terms of strain and temperature
\begin{equation}
    dS = \left(\frac{\partial S}{\partial \varepsilon}\right)_T d\varepsilon +
        \left(\frac{\partial S}{\partial T}\right)_\varepsilon dT.
\end{equation}
For an adiabatic system, the total change in entropy $dS=0$. Again rearranging and integrating,
\begin{equation} \label{DeltaT}
    \Delta T=\int_0^{\varepsilon_0} \frac{-T}{c}\left(\frac{\partial s}{\partial \varepsilon}\right)_T d\varepsilon
\end{equation}
where $c\equiv T(\partial s/\partial T)_\epsilon$ is the specific heat at constant strain. The partial derivative in the integrand is effectively the ``mechanotropic'' term. As it relates to the present work, the application of strain at constant temperature will increase the order of the microstructure as the initially amorphous material becomes nematic -- i.e., strain reduces the configurational entropy. The $(\partial s/\partial \epsilon)$ term is therefore negative, making the integrand positive. That is, the adiabatic application of strain to the present ``mechanotropic'' materials will increase their temperature.

\pagebreak
\bibliographystyle{ieeetr}
\bibliography{_bib}

\pagebreak
\renewcommand{\thefigure}{S1}
\begin{figure}
    \centering
    \includegraphics[width=0.58\textwidth]{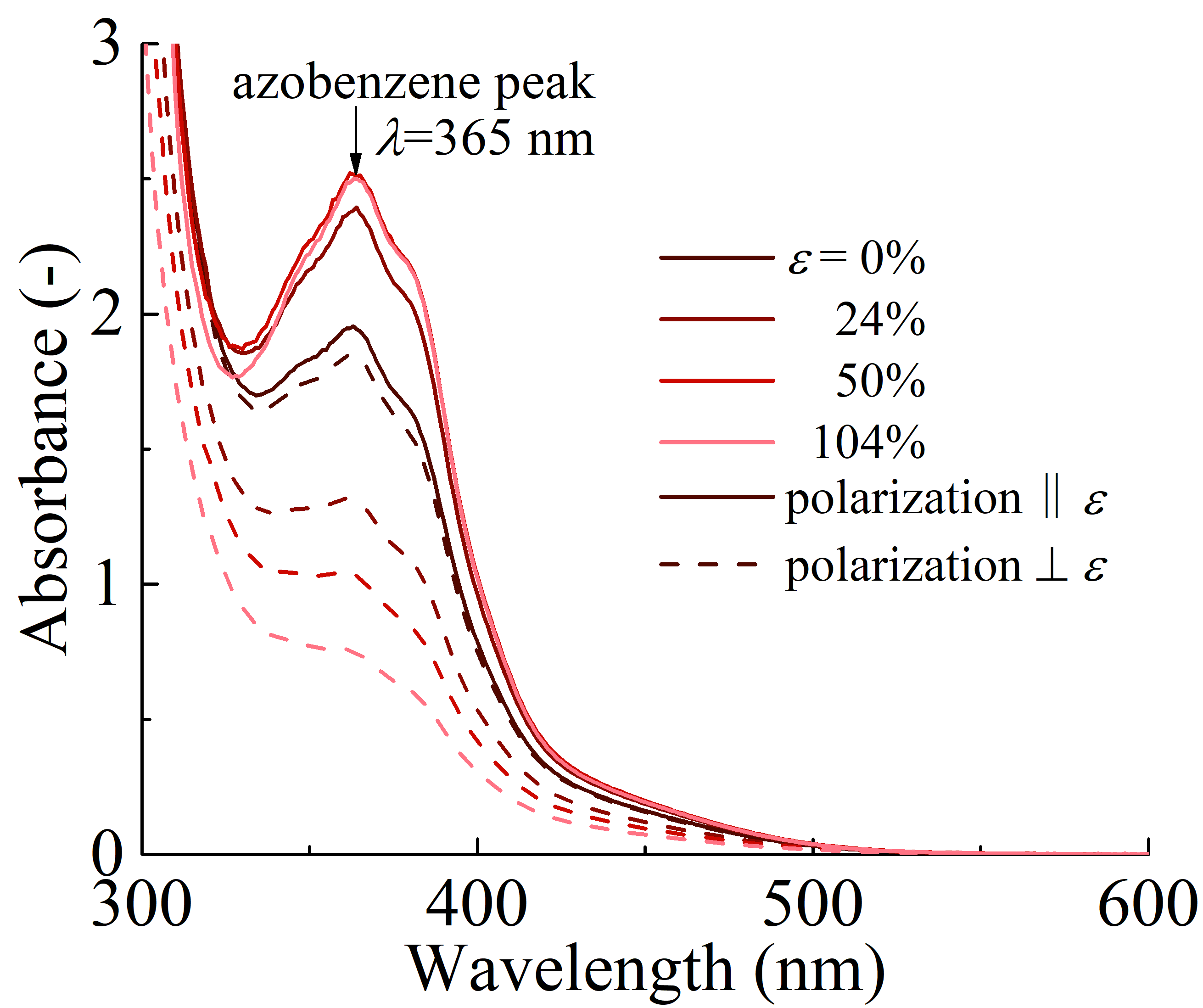}
    \caption{Typical absorbance measurements for measuring the order parameter $Q$. The dichroic azobenzene demonstrates a strong absorbance peak at $\lambda=365$~nm. Here, absorbance measurements for the material with 60~wt\% liquid crystal concentration are made at strains of $\epsilon=$0, 24, 50, and 104\%, with the incident light either polarized parallel or perpendicular to the direction of elongation. The elastomer is more absorbing of light polarized parallel to the strain, indicating the polymeric strands are orienting towards the direction of elongation. The peak absorbance for the parallel and perpendicular cases is used to calculate the order parameter $Q$ for each applied strain according to Eq.~\eqref{GHeqn}.}
    \label{fig:abs}
\end{figure}
\renewcommand{\thefigure}{S2}
\begin{figure}
    \centering
    \includegraphics[width=0.58\textwidth]{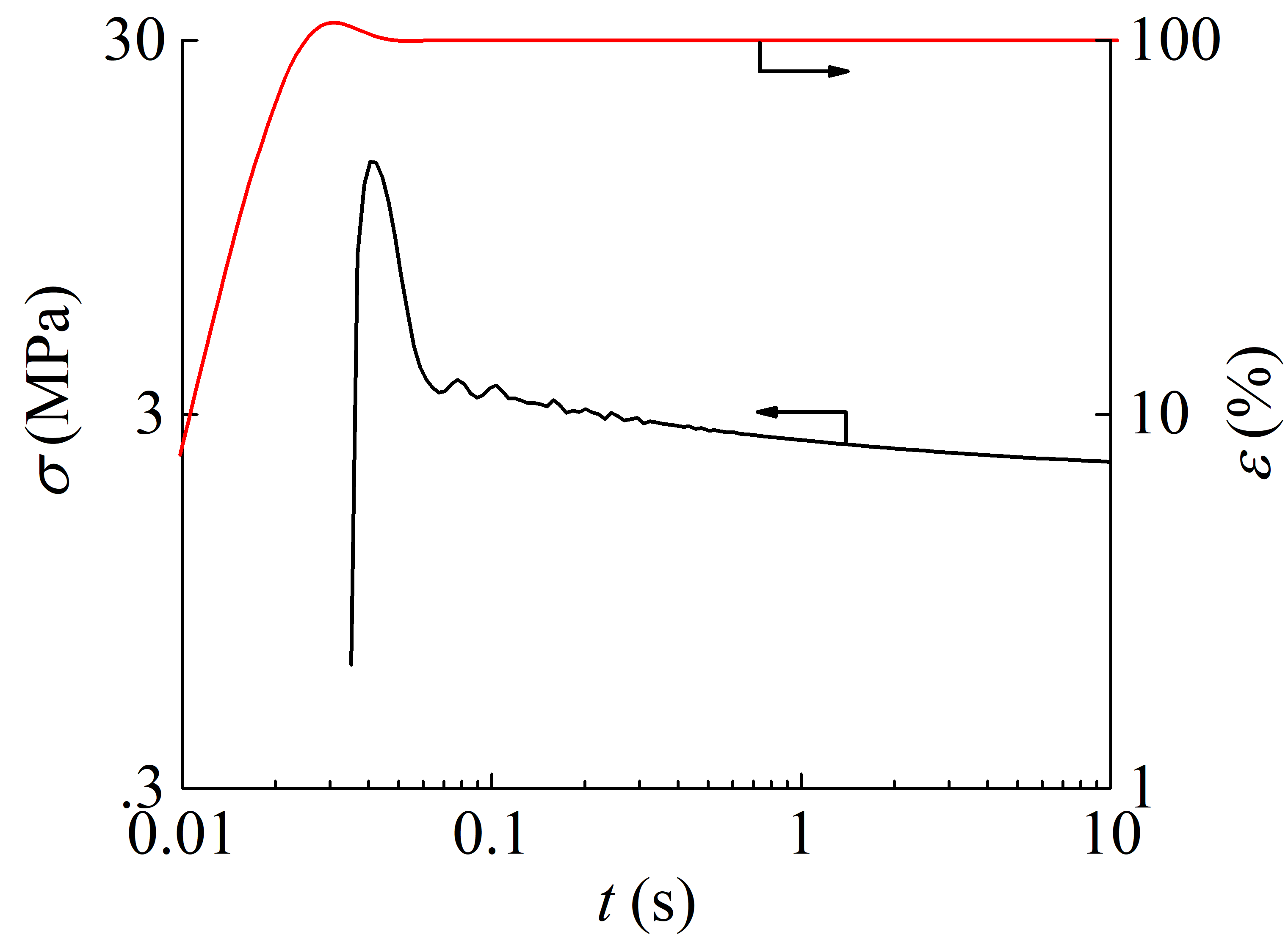}
    \caption{In response to a step-strain input (red), the stress response (black) typically demonstrates an overshoot followed by a stress decay lasting several seconds. This transient response is likely attributable to the nature of the polyacrylate polymer network.}
    \label{fig:stress}
\end{figure}